# Who Benefits, and When? Trajectories of Academic Achievement and Interest in AI-Supported Learning

Hyejeong Lee
Department of Interdisciplinary Learning and Teaching
The University of Texas at San Antonio
hyejeong.lee@utsa.edu

**Abstract:** AI-supported learning environments are increasingly used, yet relatively little is known about how students' achievement and interest develop over time or whether these trajectories vary across learners. This study examined longitudinal trajectories of mathematics achievement and interest in a semester-long AI-supported mathematics learning. Using a mixed-methods longitudinal design, repeated measures of mathematics achievement were collected at four time points and mathematics interest at three time points. Linear mixed models (LMMs) were used to examine overall changes over time, while linear quantile mixed models (LQMMs) were used to examine differences in trajectories across the lower, median, and upper portions of the outcome distributions. Mathematics achievement increased substantially during the intervention, with the largest gains emerging between the second and third measurement occasions. However, the trajectory differed across the achievement distribution. Mathematics interest increased significantly after the first month, but this increase was not sustained through the end of the semester. Qualitative findings further indicated that students experienced the AI platform differently. The study highlights the importance of examining both cognitive and motivational trajectories.

**Keywords:** Artificial intelligence; AI-supported learning; learning trajectories; achievement and interest; quantile mixed models; longitudinal mathematics learning

**Practitioner Notes**

What is already known about this topic?

- AI-supported and adaptive learning environments can provide personalized mathematics practice and immediate feedback.
- Previous research has primarily examined average changes in achievement, providing limited insight into how students at different performance levels develop over time.
- Interactive and adaptive technologies may initially increase students' interest and engagement, but the sustainability of these motivational responses remains less understood.

What this paper adds

- Students' mathematics achievement increased substantially during a semester-long AI-supported learning intervention, but growth trajectories differed across the achievement distribution.
- Higher-achieving students showed earlier gains, whereas lower- and median-achieving students demonstrated delayed improvement.
- Mathematics interest increased during the early stage of the intervention but was not sustained through the end of the semester, particularly among students with initially lower interest.
- Qualitative evidence illustrates how students experienced the same AI platform differently, including adaptive confidence-building, gamified engagement, advanced learning opportunities, flexible practice, and barriers to participation.

Implications for practice and policy

- Teachers should not assume that AI-supported learning produces the same learning trajectory for all students.
- AI-supported mathematics instruction may benefit from differentiated teacher support, particularly for students who begin with lower achievement or lower interest.
- Sustaining students' interest may require complementary teacher-led activities, meaningful challenges, and motivational support beyond the initial novelty of an AI-supported platform.
- Classroom implementation should include onboarding and technological support for students who have limited experience with digital learning environments.

## 1. Introduction

Artificial intelligence (AI) is increasingly being integrated into K-12 education where adaptive systems can provide personalized problem-solving experiences (Lee et al., 2025; Walkington & Bernacki, 2019; Yan et al., 2026). These systems are often designed to support personalized learning by dynamically adjusting task difficulty, sequencing instructional content, and providing immediate feedback based on students' performance (Holmes & Tuomi, 2022). Through such mechanisms, AI-supported learning environments are widely regarded as promising tools for improving academic outcomes and addressing diverse learner needs.

Despite this growing adoption, empirical research has largely evaluated AI-based learning through average treatment effects, focusing on whether these systems improve achievement relative to traditional instruction (Steenbergen-Hu & Cooper, 2013). Although this body of work provides evidence of positive effects on mathematics performance, it offers limited insight into how learning develops over time or whether these benefits are distributed evenly across students. In particular, relatively little is known about heterogeneous learning trajectories, that is, how students at different performance levels experience and benefit from AI-based instruction over time.

Addressing such heterogeneity is especially critical in mathematics education, where learning is inherently cumulative and strongly shaped by prior knowledge. Students enter learning environments with varying levels of preparation, which may fundamentally shape how they engage with adaptive systems. Consequently, AI-supported instruction may not function uniformly across learners. Instead, it may amplify, reduce, or reshape existing differences. However, the predominance of mean-based analytic approaches obscures these distributional dynamics, masking important variations in how students respond to AI-supported learning.

In addition to cognitive outcomes, students' motivation and interest play a critical role in shaping learning trajectories. Interest has been consistently linked to engagement, persistence, and long-term achievement in student learning. AI-based learning environments often incorporate interactive and gamified elements that may enhance students' situational interest (Acheampong et al., 2025; Hernandez et al., 2026; Walkington & Bernacki, 2019). Yet, it remains unclear whether such environments foster sustained interest over time or merely produce short-term engagement. Prior research has rarely examined the longitudinal development of both achievement and interest simultaneously, particularly within AI-supported contexts.

To address these gaps, the present study investigates students' trajectories of mathematics achievement and interest within an AI-supported learning intervention. Moving beyond mean-based approaches, this study adopts a distributional perspective to examine not only overall trends but also variation across students. In addition, qualitative data from classroom observations and student interviews are incorporated to illuminate how learners interact with and respond to the AI-based platform.

By integrating longitudinal modeling with a mixed-methods approach, this study aims to contribute to the literature in three ways. First, it examines the temporal dynamics of both achievement and interest in AI-supported learning. Second, it identifies heterogeneous learning trajectories across different performance and interest levels using a distributional perspective. Third, it provides qualitative insights into students' interaction patterns, offering a richer account of the mechanisms underlying observed quantitative trends.

## 2. Literature Review

### 2.1. AI-Based Learning in Education

Recent advances in AI have led to the increasing integration of AI-based learning platforms in educational contexts. These systems commonly incorporate adaptive learning algorithms, automated feedback, and real-time learning analytics to personalize instruction based on students' learning progress and performance patterns (du Boulay, 2019; Holmes & Tuomi, 2022; Luckin et al., 2016). By dynamically adjusting the difficulty, pacing, and sequencing of learning tasks, AI-supported environments aim to optimize practice opportunities and provide individualized support that is often difficult to achieve in traditional classroom settings (Holmes et al., 2018).

A growing body of research suggests that AI-based and adaptive learning systems can positively influence students' academic achievement. Personalized practice and immediate feedback allow learners to engage with tasks that are aligned with their current level of understanding, which can facilitate skill development and reduce learning gaps (Graesser et al., 2005; Walkington & Bernacki, 2019). In addition to cognitive outcomes, AI-supported environments may also influence motivational processes such as engagement, persistence, and interest (Ali, 2024; D'Mello & Graesser, 2013; Zhang & Aslan, 2021). Features such as interactive interfaces, self-paced learning, and adaptive scaffolding can support students' sense of competence and autonomy during learning activities.

Despite these potential benefits, empirical findings regarding the effectiveness of AI-based learning systems remain mixed. While some studies report substantial improvements in mathematics achievement (Graesser et al., 2005; Huang et al., 2016), others indicate that the magnitude of these effects varies depending on contextual and learner-related factors (Lee & Brush, 2025). For example, prior meta-analyses suggest that technology-supported learning interventions may benefit higher-performing students more than lower-performing students (Steenbergen-Hu & Cooper, 2013). Differences in implementation contexts, prior knowledge, and learner agency in AI-supported learning may contribute to these inconsistent findings (Cheung & Slavin, 2013; Hadar-Shoval, 2025; Lee et al., 2026; Steenbergen-Hu & Cooper, 2013).

Importantly, the impact of AI-based learning may evolve over time as students become more familiar with the platform or encounter increasing levels of task difficulty. Consequently, examining how learning outcomes develop across multiple time points is essential for understanding the educational impact of AI-supported instruction.

## 2.2. Heterogeneous Learning Effects and the Distributional Perspectives

Learning outcomes do not change in the same way for all students. Educational interventions often produce heterogeneous effects because learners differ in prior knowledge, learning strategies, motivation, and other individual characteristics (Cronbach & Snow, 1981). The concept of aptitude-treatment interaction (ATI) highlights that instructional approaches may be more or less effective depending on the characteristics of individual learners (Snow, 1991). Although ATI effects are often complex and difficult to identify in practice (Snow, 2013), the framework underscores the importance of considering individual differences when evaluating instructional interventions.

Students typically enter learning environments with varying levels of prior knowledge and skills, which strongly influence subsequent learning outcomes (Van Kesteren et al., 2014). In mathematics education, such differences are particularly salient because mathematical knowledge is cumulative and hierarchical in nature. Students with higher initial proficiency is a strong predictor of subsequent learning (Booth et al., 2014; Wilkins & Norton, 2018). Conversely, students with lower prior achievement may require additional time to adapt to new learning, leading to accumulated disadvantages and lagging behind (Fuchs et al., 2012).

These differences can lead to distinct developmental trajectories across the lower, middle, and upper portions of the achievement distribution. However, many studies of educational interventions rely primarily on mean-based analyses, which estimate average changes across all students. While such analyses provide useful summaries of overall trends, they may obscure important variations in how different groups of students experience instructional interventions.

A distributional perspective offers an alternative approach by examining changes across different parts of the outcome distribution. Linear quantile mixed models (LQMM) extend conventional mixed-effects models by estimating conditional quantiles rather than conditional means while accounting for the hierarchical structure of repeated-measures data. This approach enables researchers to investigate whether lower-, median-, and higher-performing students exhibit different trajectories over time.

Such an approach is particularly relevant in AI-based learning environments, where adaptive systems dynamically respond to individual performance and learning behaviors. As a result, the effects of AI-supported instruction may vary substantially depending on students' starting points and interaction patterns with the learning platform.

## 2.3. Trajectories of Mathematics Achievement and Interest Over Time

Learning outcomes in educational interventions often unfold gradually rather than appearing immediately. For this reason, longitudinal perspectives have become increasingly important for understanding how students' learning and motivational experiences develop over time. In mathematics education, repeated-measures designs have been widely used to examine developmental trajectories of both academic achievement and affective outcomes.

Mathematics achievement may increase as students accumulate practice opportunities and receive feedback that helps them address misconceptions and refine problem-solving strategies. Technology-supported learning environments, particularly adaptive systems, can facilitate such learning processes by providing individualized learning pathways (Holmes et al., 2018; Walkington & Bernacki, 2019). However, improvements in achievement may not occur uniformly throughout an instructional intervention. Early gains may reflect initial engagement, whereas later stages may involve slower progress as tasks become more cognitively demanding.

In addition to academic achievement, affective factors such as motivation and interest may also have different patterns and changes. Digital learning environments, including AI-based platforms, have been suggested to foster interest through AI-based mechanisms (Acheampong et al., 2025; Walkington & Bernacki, 2019; Yeh et al., 2019). Interactive interfaces, immediate feedback, and opportunities for self-paced learning can create engaging learning and fun experiences that differ from traditional teacher-directed instruction (Awang et al., 2025). Situational interest may be triggered by engaging learning environments, while sustained individual interest emerges when learners repeatedly experience meaningful engagement and growing competence (Köller et al., 2001; Yeh et al., 2019). However, maintaining interest over time may require sustained perceptions of competence and meaningful challenge.

Students who perceive adaptive feedback as supportive may experience increased engagement, whereas others may view frequent automated feedback or practice requirements as monotonous or demanding. These differences suggest that students' achievement and interest trajectories in AI-supported mathematics learning may not follow a uniform pattern across the instructional interventions. Understanding how students' achievement and interest evolve over time in AI-based learning contexts is therefore an important area of inquiry. Examining trajectories of mathematics achievement and interest across multiple time points can provide a more comprehensive understanding and insights of students' learning experiences in AI-based mathematics learning environments whether initial engagement with AI-supported learning environments is maintained, strengthened, or diminished throughout an instructional intervention.

### 2.4. Research Gap and Rationale for the Present Study

Although prior research has examined the effects of technology-supported mathematics learning on students' achievement and motivation, several limitations remain. First, many studies rely on pre-post designs that capture only two time points, making it difficult to understand how learning outcomes evolve throughout an intervention. Such designs may overlook important temporal patterns, including early increases followed by stabilization or decline (Lee, 2025).

Second, much of the existing literature focuses on average treatment effects, which may obscure heterogeneous responses among students. In adaptive learning environments, instructional experiences are shaped by individual performance and interaction with the system, suggesting that students at different points of the outcome distribution may experience distinct developmental trajectories. For example, students with lower initial achievement may require time to adapt to the platform before demonstrating improvement, whereas higher-performing students may benefit more quickly from advanced problem-solving opportunities.

Third, while mathematics achievement has received considerable attention in studies of technology-supported learning, relatively fewer studies have examined how students' interest in mathematics evolves over time within AI-based learning environments. Even when motivational outcomes are included, analyses often focus on overall changes rather than examining whether interest trajectories differ across groups of students.

To address these limitations, the present study investigates both mathematics achievement and mathematics interest trajectories in an AI-based mathematics learning environment using repeated measures data. In addition to examining overall changes over time, the study adopts a distributional perspective by applying quantile models to estimate trajectories at the lower, median, and upper portions of the outcome distribution. This approach allows for a more nuanced understanding of how students with different performance levels experience AI-supported mathematics learning.

### 2.5. Research Questions

Based on the theoretical and empirical considerations discussed above, the present study addresses the following research questions:

RQ1. How do students' mathematics scores change across measurement time points during AI-based mathematics learning, and do these trajectories differ across the lower, median, and upper portions of the score distribution?

RQ2. How does students' mathematics interest change across measurement time points during AI-based mathematics learning, and do these trajectories differ across the lower, median, and upper portions of the interest distribution?

RQ3. How do students interact with the AI-based mathematics platform during classroom learning activities, and how do these interactions shape their learning experiences and engagement with the system?

## 3. Method

### 3.1. Research Design

This study employs mixed-methods design. (Creswell & Clark, 2017; Creswell & Creswell, 2017) to investigate students' learning and motivational trajectories in an AI-supported mathematics classroom. This mixed-methods approach is advantageous in triangulation, complementarity, and expansion of the data interpretation to complement each other (Greene et al., 1989). In the present study, quantitative data were used to examine longitudinal changes in students' mathematics achievement and interest, while qualitative data were used to explore how students interacted with the AI-based learning environment.

The rationale for adopting a mixed-methods design lies in the complementary strengths of the two approaches. While quantitative analyses allow for the identification of overall trends and distributional patterns, they provide limited insight into the mechanisms underlying these patterns. Qualitative data, including classroom observations, field notes, and student interviews, were therefore incorporated to contextualize and explain the observed trajectories. This design enabled a more nuanced understanding of both the effectiveness of the AI-based intervention and students' personalized learning experiences.

### 3.2. Participants, AI Platform, and Procedure

A total of twenty-three third-grade students from a public elementary school participated in this study. The participating school served students from low socioeconomic backgrounds, including students from immigrant and single-parent families. This context is particularly relevant for examining the potential of AI-supported learning to address educational inequities. The intervention was conducted over one academic semester within a third-grade mathematics classroom.

Students used an AI-based mathematics learning platform, *Tok-Tok Math Expedition*. The platform integrates several instructional features, including digital textbooks, AI-recommended problem sets, interactive games, and automated feedback. A key feature of the platform is its adaptive recommendation system, which analyzes students' responses and learning histories to provide personalized problem sequences. The system adjusts task difficulty in real time and provides immediate feedback to support students' learning processes.

Students' mathematics achievement was assessed at four time points: (1) beginning of the semester, (2) after one month, (3) after two months, and (4) at the end of the semester (approximately four months). Students' mathematics interest was measured at three time points: (1) beginning of the semester, (2) after one month, and (3) at the end of the semester.

### 3.3. Data Sources

The data sources used for this study are mathematics achievement, mathematics interest, classroom observations, field notes, and informal student interviews. The math achievement assessments were developed through a collaborative process involving researchers and mathematics education experts. The development process involved iterative review and revision to ensure content validity. Pilot testing with four students was conducted to refine item clarity and difficulty. Each assessment consisted of 20 items, with each item worth five points, resulting in a maximum score of 100 points per assessment.

Students' mathematics interest was measured using items adapted from established motivational instruments, mathematics motivation questionnaire (Fiorella et al., 2021) and mathematics interest items (Frenzel et al., 2010). Items were adapted to be developmentally appropriate for third-grade students. Responses were recorded on a five-point Likert scale ranging from 1 (strongly disagree) to 5 (strongly agree). Internal consistency reliability was assessed using Cronbach's alpha. The reliability coefficients were high across administrations ($\alpha = .882$ at pretest; $\alpha = .891$ at posttest), indicating acceptable internal consistency.

Qualitative data were collected through classroom observations, field notes, and informal student interviews conducted during and after AI-based learning activities. Observations focused on students' interaction patterns with the AI platform, engagement behaviors, and notable verbal responses. Field notes were recorded throughout the intervention, and illustrative student quotes were documented when meaningful behaviors or experiences emerged. Informal interviews were conducted to capture students' perceptions of the AI-based learning environment and their learning experiences.

### 3.4. Data Analyses

The collected qualitative data were analyzed using thematic analysis following the procedures outlined by Braun and Clarke (2006). The rationale for adopting a mixed-methods design lies in the complementary strengths of the two approaches. While quantitative analyses allow for the identification of overall trends and distributional patterns, they provide limited insight into the mechanisms underlying these patterns. Qualitative data, including classroom observations, field notes, and student interviews, were therefore incorporated to contextualize and explain the observed trajectories. This design enabled a more nuanced understanding of both the effectiveness of the AI-based intervention and students' personalized learning experiences.

For the quantitative data, we employed a two-step analytical approach comprising Linear Mixed Models (LMM) and Linear Quantile Mixed Models (LQMM) to investigate the longitudinal effects of the AI-based instruction on students' math achievement and interest. All statistical analyses were performed using the R statistical computing environment.

First, LMM was fitted to examine the average trajectories of students' math scores and interest levels over time. LMM is particularly suitable for longitudinal data as it accounts for the nested structure of repeated measures within individuals and handles missing data effectively under the missing-at-random assumption (Laird & Ware, 1982; Pinheiro & Bates, 2000). We specified the model with time as a fixed effect to estimate the overall growth trend, and included random intercepts for individual students to account for between-student variability in baseline levels (Bates et al., 2015).

Second, to explore potential heterogeneous effects of the intervention across different math proficiency and interest levels, we conducted LQMM. While LMM estimates the effects on the conditional mean of the outcome distribution, LQMM allows for the estimation of effects at specific quantiles (e.g., 25th, 50th, and 75th percentiles), thereby providing a more comprehensive picture of the distributional changes (Geraci & Bottai, 2014). This approach is advantageous for identifying whether the intervention benefits low-performing (or low-interest) students differently compared to their high-performing counterparts. We utilized the lqmm package in R, estimating the parameters based on the asymmetric Laplace distribution (Geraci & Bottai, 2007; Geraci, 2014).

## 4. Results

### 4.1. Math Performance Trajectory

To examine the first research question, mathematics scores across the four measurement occasions were measured. Table 1 presents the descriptive statistics for math achievement scores. At Time 1, the mean score was 54.2 (median = 55.0), with relatively large variability (SD = 19.3; min = 20.0, max = 85.0). At Time 2, both the mean (57.3) and median (60.0) increased slightly; however, score dispersion increased substantially (SD = 27.7; min = 20.0, max = 100.0). Notably, the 25th percentile decreased from 45.0 at Time 1 to 30.0 at Time 2, suggesting that lower-performing students may not have improved at the same rate as their peers during this interval. A marked improvement was observed at Time 3, where the mean rose to 79.5 and the median to 82.5. Variability also decreased considerably (SD = 10.8), indicating a more concentrated distribution of scores (25th percentile = 77.5; 75th percentile = 85.0; min = 47.5, max = 95.0). At Time 4, the mean (73.3) and median (75.0) declined relative to Time 3 but remained higher than those at Times 1 and 2 (min = 50.0, max = 100.0). The standard deviation increased again (SD = 17.7), suggesting that individual differences widened at the final measurement.

**Table 1**
*Descriptive Statistics of Math Achievement Scores*

| Time | Mean | Median | SD | Min | Max | 25th | 50th | 75th |
|---|---|---|---|---|---|---|---|---|
| 1 | 54.2 | 55.0 | 19.3 | 20.0 | 85.0 | 45.0 | 55.0 | 70.0 |
| 2 | 57.3 | 60.0 | 27.7 | 20.0 | 100.0 | 30.0 | 60.0 | 80.0 |
| 3 | 79.5 | 82.5 | 10.8 | 47.5 | 95.0 | 77.5 | 82.5 | 85.0 |
| 4 | 73.3 | 75.0 | 17.7 | 50.0 | 100.0 | 56.3 | 75.0 | 85.0 |

The scatter plot in Figure 1 corroborates these trends, showing an overall rise in performance from Time 1 to Time 3, followed by a modest decrease at Time 4. The Time 3 distribution appears most compact, consistent with the reduced variability observed in the descriptive statistics.

**Figure 1**

*Repeated Math Performance Scatter Plot*

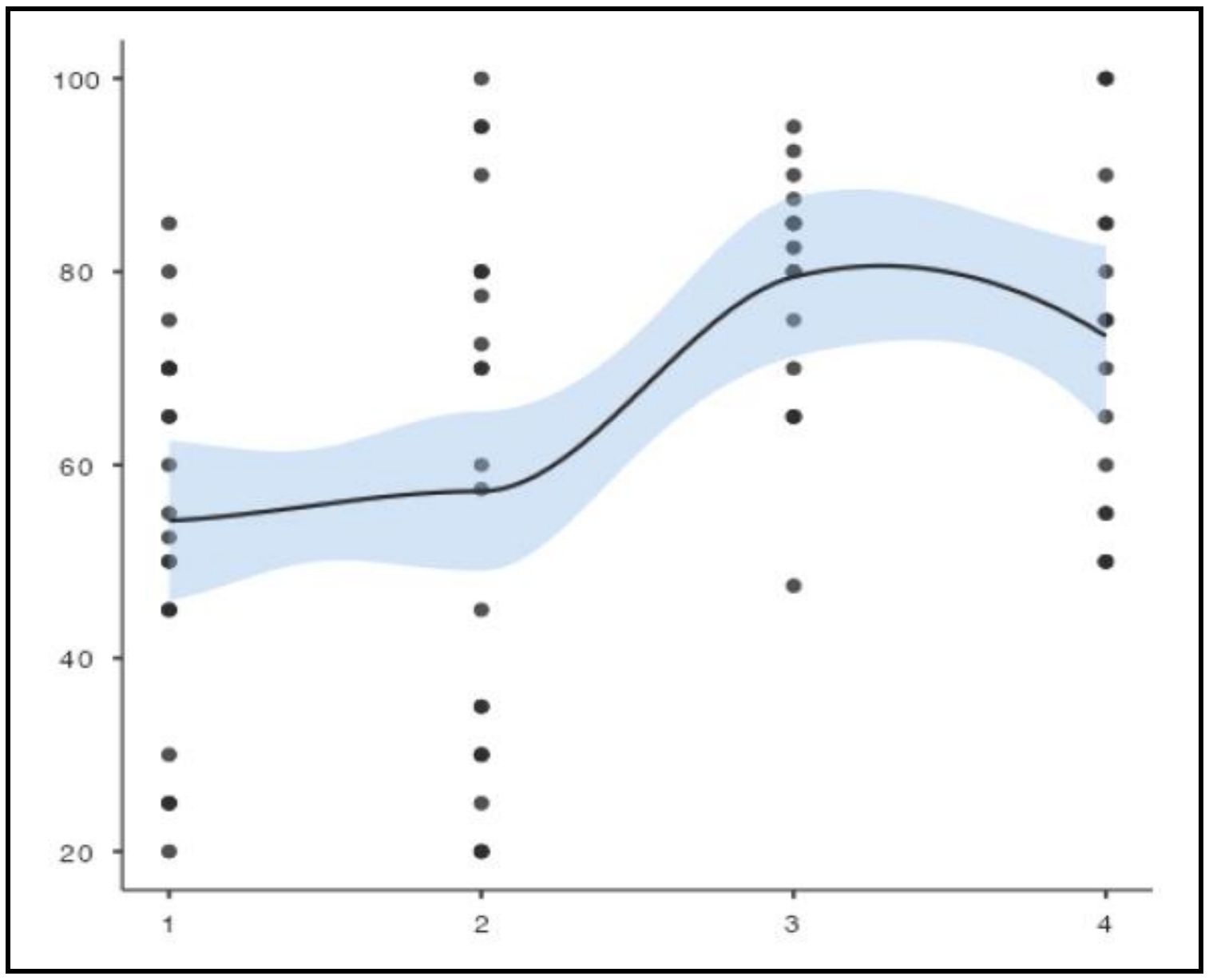

To test whether these descriptive changes in math achievement scores across four time points were statistically significant, we fitted the LMM, specifying time as a fixed effect and including a random intercept for students to account for repeated measures (see Table 2). The random-effects results indicated substantial between-student variability in baseline performance ($\sigma^2_{ID}$ = 182), indicating significant individual differences in baseline math achievement. The residual variance was 213.0, reflecting within-student variability over time. The intraclass correlation coefficient (ICC) was .460, suggesting that approximately 46.0% of the total variance in math scores was attributable to stable differences between students. This substantial proportion of variance at the student level justifies the use of a linear mixed model.

The model explained 23.6% of the variance through fixed effects (marginal $R^2$ = .236), and the full model including both fixed and random effects accounted for 58.7% of the variance (conditional $R^2$ = .587). The fixed effect of time was statistically significant, indicating that math performance changed across measurement occasions, $F(3, 61.7) = 16.3$, $p < .001$.

Table 2 presents the fixed effect parameter estimates (with Time 1 as the reference). Compared to Time 1, the change at Time 2 was not statistically significant (B = 3.04, SE = 4.31), 95% CI [-5.52, 11.60], t(61.1) = .71, $p = .482$. In contrast, scores were significantly higher at Time 3 compared with Time 1 (B = 25.22, SE = 4.31), 95% CI [16.65, 33.80], t(61.1) = 5.86, $p < .001$, and also significantly higher at Time 4 compared with Time 1 (B = 20.49, SE = 4.66), 95% CI [11.22, 29.80], t(62.4) = 4.40, $p < .001$.

**Table 2**

*LMM Estimates for Fixed and Random Effects for Math Achievement over Time*

| Fixed Effects | Estimate | SE | 95% CI Lower | 95% CI Upper | df | t | *p* |
|---|---|---|---|---|---|---|---|
| Intercept | 66.43 | 3.23 | 60.01 | 72.8 | 22.2 | 20.587 | <.001 |
| Time 2 | 3.04 | 4.31 | -5.52 | 11.6 | 61.1 | .707 | .482 |
| Time 3 | 25.22 | 4.31 | 16.65 | 33.8 | 61.1 | 5.856 | <.001 |
| Time 4 | 20.49 | 4.66 | 11.22 | 29.8 | 62.4 | 4.395 | <.001 |
| Random Effects | Var | SD | ICC | | | | |
| Intercept (ID) | 182 | 13.5 | .460 | | | | |
| Residual | 213 | 14.6 | | | | | |

Post hoc pairwise comparisons with Bonferroni adjustment indicated that the difference between Time 1 and Time 2 was not statistically significant (Δ = 3.04, SE = 4.31, t = 0.71, *p* = 1.000). In contrast, scores at Time 3 were significantly higher than at Time 1 and Time 2 (Time 1 vs. Time 3: Δ = 25.22, SE = 4.31, t = 5.86, *p* < .001; Time 2 vs. Time 3: Δ = 22.17, SE = 4.31, t = 5.15, *p* < .001). Scores at Time 4 were also significantly higher than at Time 1 and Time 2 (Time 1 vs. Time 4: Δ = 20.49, SE = 4.66, t = 4.40, *p* < .001; Time 2 vs. Time 4: Δ = 17.45, SE = 4.66, t = 3.74, *p* = .002). Crucially, however, the comparison between Time 3 and Time 4 was not significant (Δ = 4.72, SE = 4.66, t = 1.01, *p* = 1.000). These post hoc results indicate that the rapid growth observed in Time 3 did not continue. Instead, the students' performance plateaued at the end of the semester rather than continuing to increase.

To determine whether this overall pattern obscured heterogeneity across the score distribution, LQMMs were estimated at *τ* = .25, .50, and .75. As shown in Table 3, the timing of improvement differed across quantiles. At the 25th and 50th quantiles, no significant change was observed at Time 2 relative to baseline (*ps* > .05), whereas significant gains emerged at both Time 3 and Time 4 (all *ps* < .001). In contrast, at the 75th quantile, achievement was already significantly higher at Time 2, B = 13.25, SE = 5.03, *p* = .011, and remained elevated through Time 3 and Time 4 (both *ps* < .001).

**Table 3**

*Parameter Estimates for LQMM on Math Achievement Scores*

| | τ = .25 | | | τ = .50 | | | τ = .75 | | |
|---|---|---|---|---|---|---|---|---|---|
| | Estimate | SE | *p* | Estimate | SE | *p* | Estimate | SE | *p* |
| Intercept | 43.43 | 9.9 | < .001 | 55.03 | 5.68 | < .001 | 65.06 | 3.96 | < .001 |
| Time 2 | -4.23 | 8.32 | .613 | 4.73 | 7.41 | .526 | 13.25 | 5.03 | .011 |
| Time 3 | 28.86 | 4 | < .001 | 24.97 | 3.99 | < .001 | 19.94 | 3.67 | < .001 |
| Time 4 | 19.28 | 5.02 | < .001 | 19.97 | 4.84 | < .001 | 19.89 | 3.97 | < .001 |

Figure 2 complements these estimates by illustrating quantile-specific trajectories across time. The lower and median quantiles exhibited a delayed improvement pattern, characterized by minimal change through Time 2, a pronounced increase at Time 3, and some attenuation at Time 4. The upper quantile, by contrast, showed earlier gains and a comparatively stable trajectory after Time 2. Overall, the LQMM findings indicate that the mean trajectory captured by the LMM masked important distributional differences in students' learning trajectories during the AI-supported mathematics intervention.

**Figure 2**

*Math Achievement Score Trajectory by Quantile*

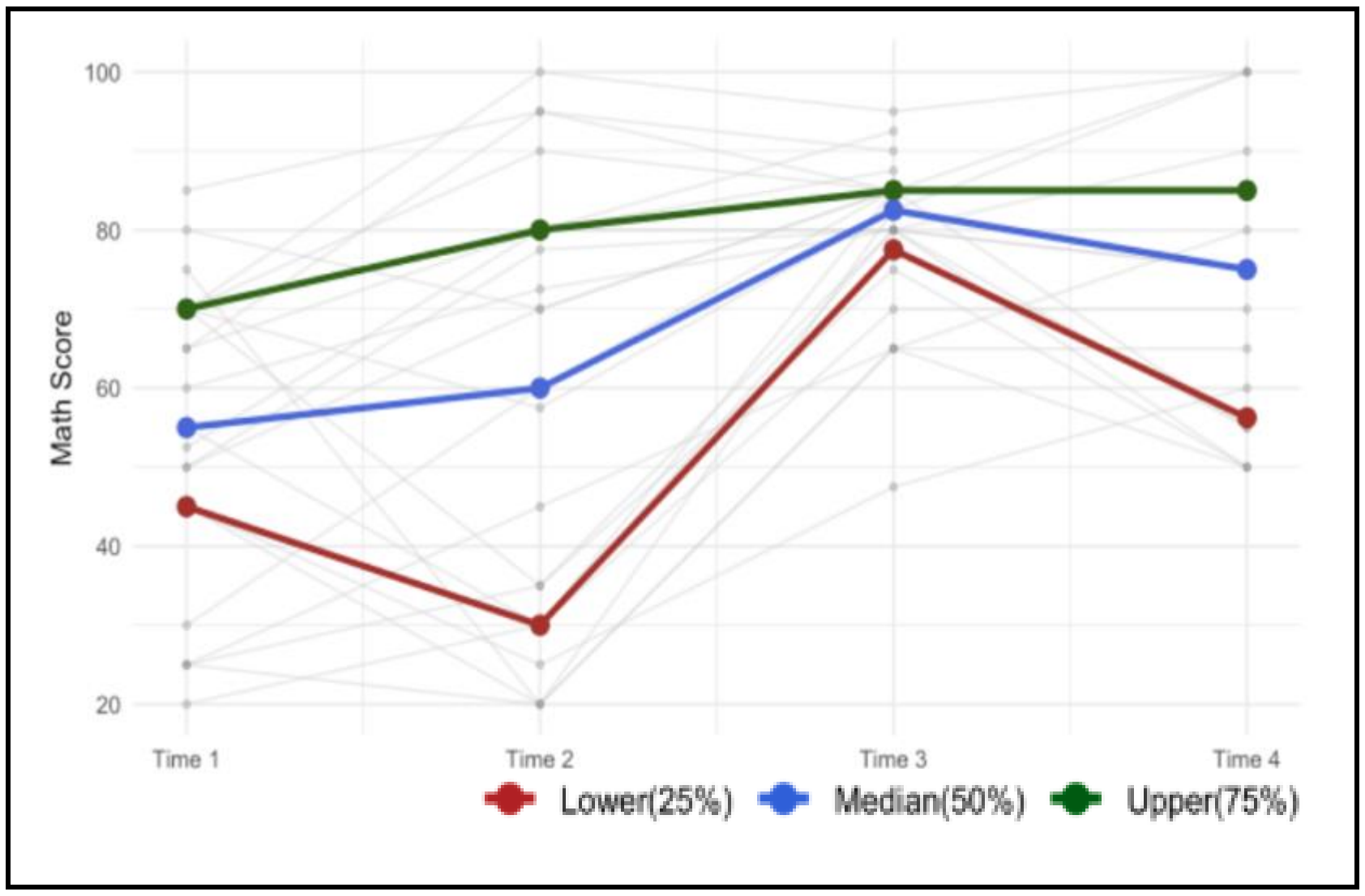

## 4.2. Math Interest Trajectory

Table 4 summarizes students' math interest across three measurement occasions (Time 1 baseline, Time 2 one month later, Time 3 end of semester, approximately four months later). At baseline, the mean interest score was 3.38 (SD = .77). Interest increased modestly after one month (M = 3.60, SD = .79). By the end of the semester, however, interest decreased to 3.46 (SD = .63) compared to Time 2. As shown in Figure 3, the overall pattern suggests a short-term uptick in interest early in the intervention that did not persist through the end of the semester, with the trend tapering back toward baseline levels.

**Table 4**

*Descriptive Statistics of Math Interest Scores*

| Time | Mean | Median | SD | Min | Max | 25th | 50th | 75th |
|---|---|---|---|---|---|---|---|---|
| 1 | 3.38 | 3.44 | 0.767 | 2.00 | 5.00 | 2.83 | 3.44 | 3.89 |
| 2 | 3.60 | 3.67 | 0.785 | 1.89 | 5.00 | 3.17 | 3.67 | 4.00 |
| 3 | 3.46 | 3.42 | 0.633 | 2.44 | 5.00 | 3.10 | 3.42 | 3.79 |

**Figure 3**

*Repeated Math Interest Scatter Plot*

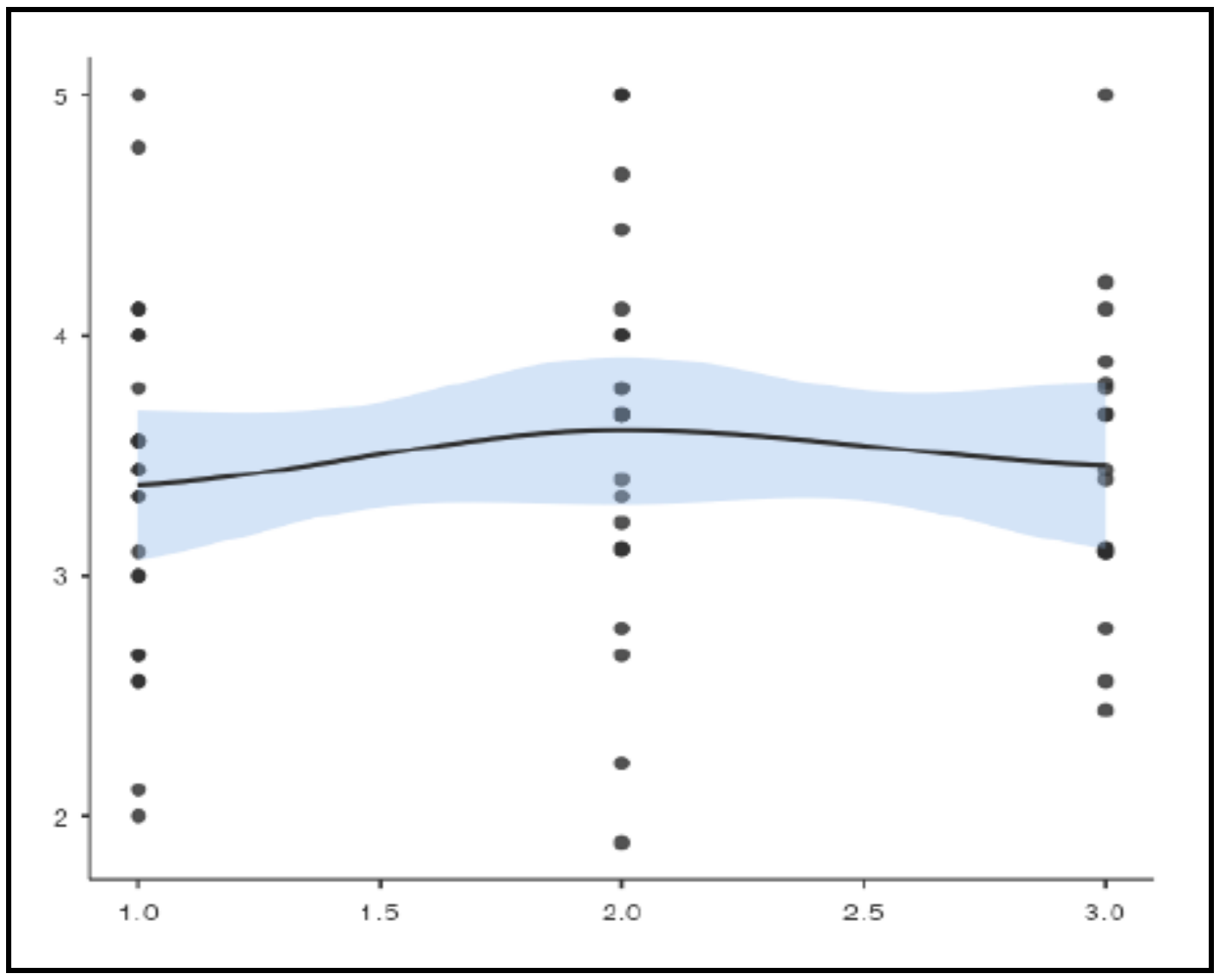

To test whether these descriptive changes in math interest were statistically significant, we fitted the LMM with time as a fixed effect and a random intercept for students. The analysis of random effects showed substantial variability in the baseline interest levels among students ($\sigma^2_{ID}$ = .43), suggesting meaningful individual differences in interest levels. The residual variance was 0.119, reflecting within-student variability across measurement occasions. The ICC was calculated as .783, indicating that approximately 78.3% of the variance in math interest was attributable to stable between-student differences. This high ICC justifies the use of a mixed effects model with a random intercept for students to account for the non-independence of the repeated measures.

To examine the longitudinal changes in students' math interest, a linear mixed model was fitted with time as a fixed effect and student ID as a random effect. The model results revealed a statistically significant increase in math interest at Time 2 compared to the baseline (B = .23, SE = .10, $p$ = .032). This indicates that students' interest significantly improved after the first month of the intervention. However, by Time 3 (approximately four months later), the increase in interest was not statistically significant compared to the baseline (B = 0.11, SE = 0.11, $p$ = .329). Although the mean interest score at Time 3 remained numerically higher than the baseline, the statistical analysis suggests that the initial boost in interest observed at Time 2 may have leveled off by the end of the semester.

**Table 5**

*LMM Estimates for Fixed and Random Effects for Math Interest over Time*

| Fixed Effects | Estimate | SE | 95% CI Lower | 95% CI Upper | df | t | *p* |
|---|---|---|---|---|---|---|---|
| Intercept | 3.489 | .144 | 3.201 | 3.777 | 22.0 | 24.274 | <.001 |
| Time 2 | .226 | .102 | .022 | .429 | 39.0 | 2.219 | .032 |
| Time 3 | .110 | .111 | -0.113 | .333 | 39.5 | .989 | .329 |
| **Random Effects** | **Var** | **SD** | **ICC** | | | | |
| Intercept (ID) | .430 | .656 | .783 | | | | |
| Residual | .119 | .345 | | | | | |

To examine whether changes in math interest scores over time differed across the outcome distribution ($\tau = .25, .50, .75$), LQMMs were conducted. The results, presented in Table 6, indicate that the trajectory of math interest varied depending on the students' initial levels of interest.

**Table 6**
*Parameter Estimates for LQMM on Math Interest Score*

| | τ = .25 | | | τ = .50 | | | τ = .75 | | |
|---|---|---|---|---|---|---|---|---|---|
| | Estimate | SE | *p* | Estimate | SE | *p* | Estimate | SE | *p* |
| Intercept | 3.03 | .21 | < .001 | 3.43 | .14 | < .001 | 3.64 | .19 | < .001 |
| Time 2 | .19 | .14 | .171 | 0.24 | .09 | 0.011 | .22 | .09 | .015 |
| Time 3 | .07 | .15 | .630 | 0.09 | .14 | 0.532 | .14 | .12 | .234 |

As shown in Figure 4, although a slight numeric increase was observed across all groups, the statistical significance varied. Students in the lower quartile of interest ($\tau = .25$) did not show statistically significant improvements at any time point. Their interest levels remained relatively low throughout the semester, with no significant changes observed at either Time 2 ($p = .171$) or Time 3 ($p = .630$) compared to the baseline. This lack of significance is likely attributable to larger standard errors, which implies greater heterogeneity in how lower-interest students responded to the intervention.

For students in the median ($\tau = .50$) and high-interest ($\tau = .75$) groups, the intervention showed a clear initial increase. Specifically, at Time 2 (one month into the intervention), the median group showed an increase in interest compared to the baseline (B = .24, $p = .011$). Similarly, the high-interest group demonstrated an increase (B = .22, $p = .015$). However, this upward trend was not sustained until the end of the semester; by Time 3, the interest scores for all groups were not statistically distinguishable from their baseline levels ($p > .05$).

Overall, the quantile specific results indicate that increases in math interest were most evident for students in the middle and upper portions of the interest distribution, while students with relatively low interest did not show reliable change across the three occasions.

**Figure 4**

*Math Interest Score Trajectory by Quantile*

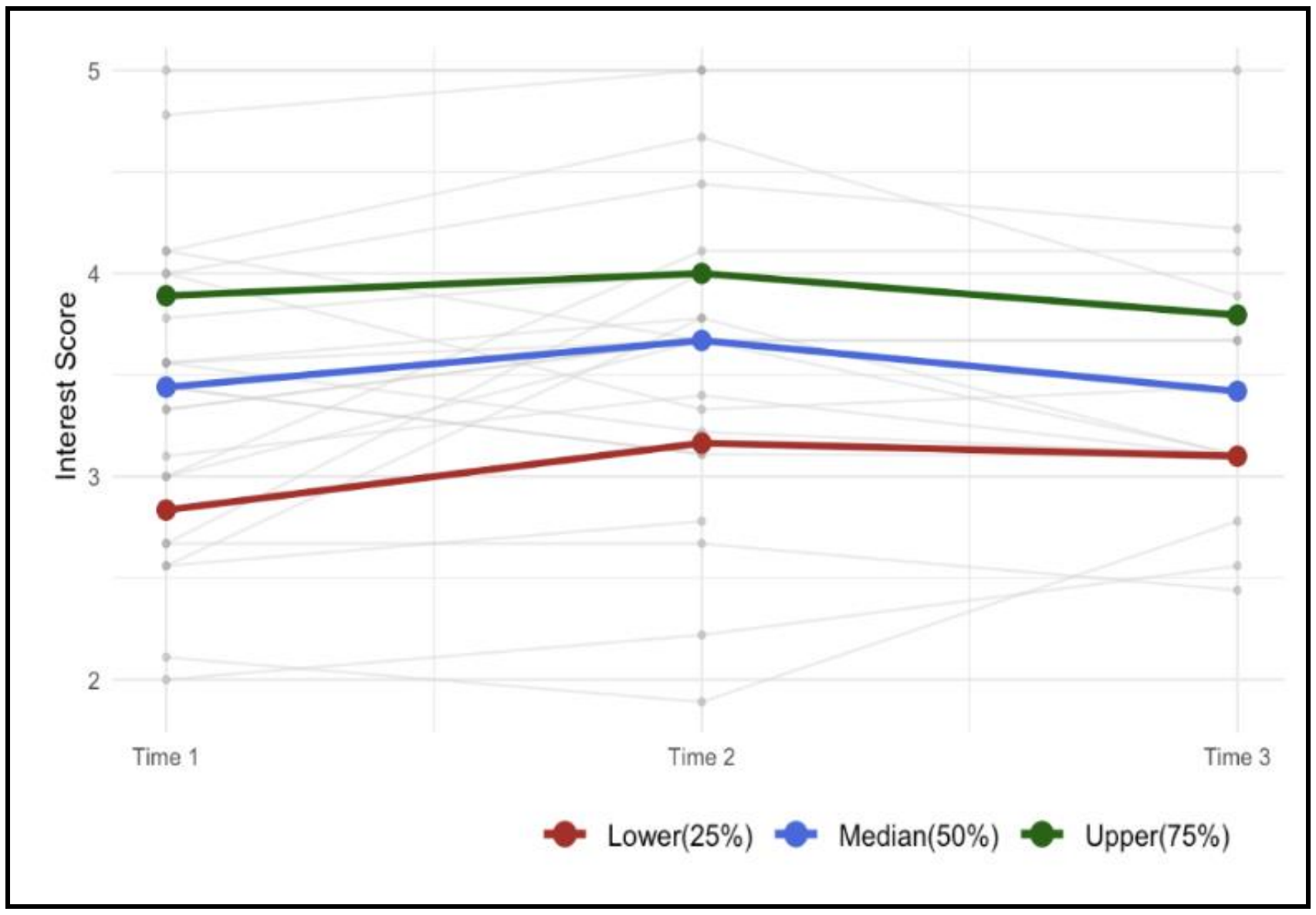

**4.3. Students' Interaction Patterns with the AI Platform**

To complement the quantitative analyses of achievement and interest trajectories, qualitative data from classroom observations, field notes, and student interviews were analyzed to examine how students interacted with the AI-based mathematics platform during classroom learning activities. The qualitative findings revealed several distinct interaction patterns (see Figure 5 & Table 7) that shaped students' learning experiences with the platform. These patterns highlight how students with different learning profiles engaged with the system in different ways.

**Figure 5**

*Patterns of Student Interaction with the AI Platform*

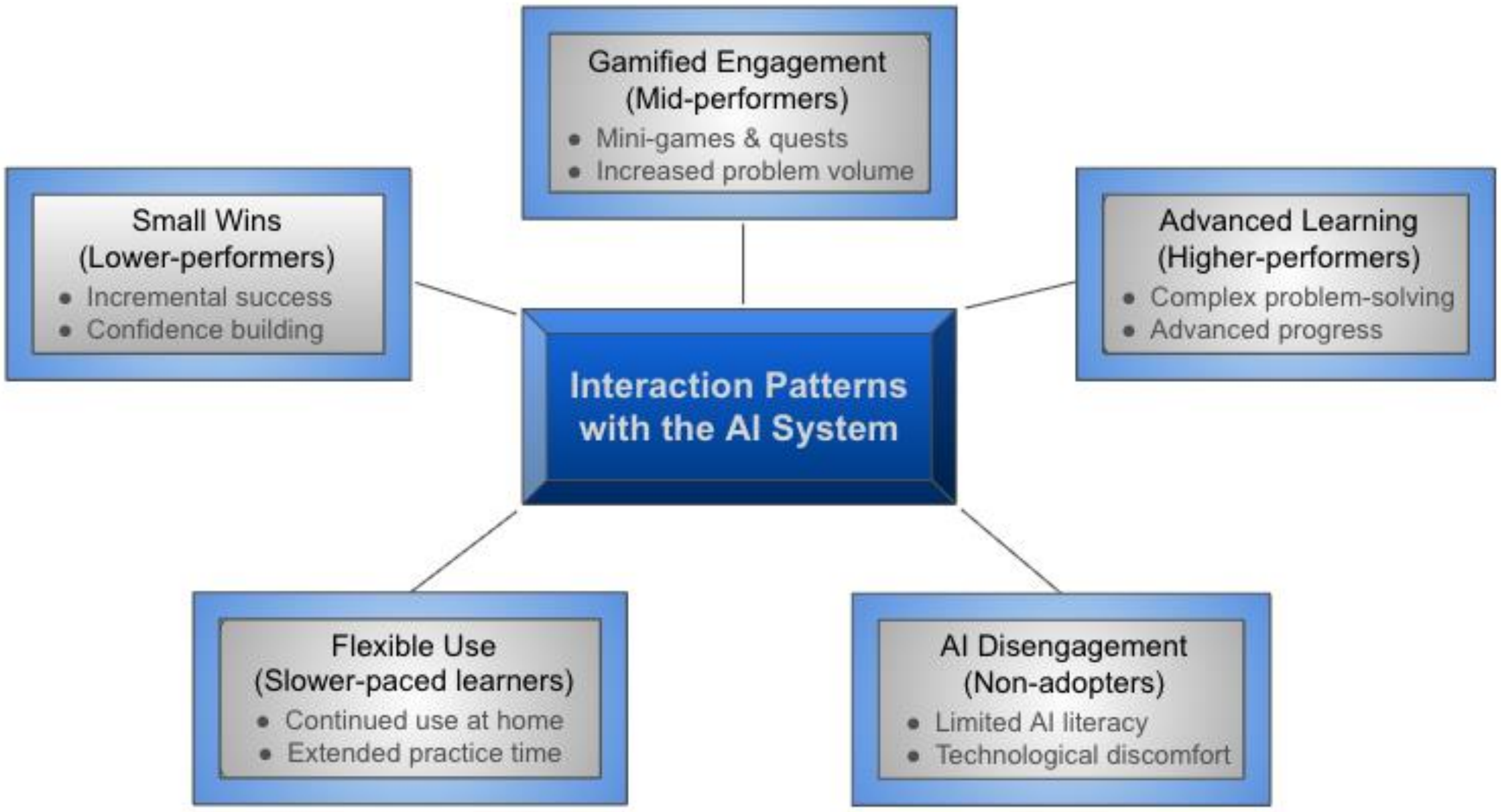

**Table 7**
*Qualitative Themes by Learner Profile*

| Learner Profile | Theme | Key Indicators |
|---|---|---|
| Lower-achieving learners | Confidence Through Small Wins | Repeated success with simple problems; emerging sense of accomplishment; increased confidence |
| Mid-achieving learners | Gamified Engagement Through Repeated Practice | Mini-games and quests; increased problem volume; sustained practice |
| Higher-achieving learners | Seeking Greater Cognitive Challenge | Complex problem-solving; advanced progress; movement beyond basic tasks |
| Slower-paced learners | Flexible, Extended Use Across Contexts | Continued use at home; use during breaks; extended practice time |
| Non-adopters | AI Disengagement | Limited AI literacy; technological discomfort; low participation |

### *4.3.1. Confidence through Small Wins*

For some lower-achieving students, the adaptive nature of the AI platform created opportunities to experience early success, which contributed to increased confidence in mathematics. Because the system recommended problems aligned with students' current ability levels, learners who previously struggled with grade-level content were able to begin with easier tasks and gradually progress to more advanced problems.

Rachael, a student who had recently transferred schools and experienced gaps in prior learning, demonstrated substantial improvement during the intervention. The adaptive recommendation system allowed her to practice foundational problems before progressing to grade-level tasks. During the interview, she described how these experiences contributed to a growing sense of competence:

> *"I didn't have confidence in math. I couldn't understand what Ms. Cho said in math class. But when I used the tool, AI recommended very easy questions. When I tried third-grade math problems, I got them wrong. But when I solved first-grade problems, I kept getting them right. I was so happy. Now I can solve the third-grade math problems, and I understand what Ms. Cho explains in class. I feel confident now."*

This experience illustrates how adaptive learning systems can support struggling learners by enabling incremental mastery experiences.

#### *4.3.2. Gamified Engagement through Extensive Practice*

Several students reported that the gamified features of the AI platform encouraged them to engage in extensive mathematics practice. Mini-games, quests, and leveling systems appeared to transform repetitive problem solving into a more engaging learning activity.

Jake, a student with intermediate achievement, described how the platform motivated him to complete a large number of practice problems:

> *"I played a lot of mini-games on the AI system. I was really into it. It didn't feel like studying. I liked math class because I could play games. Math felt like a fun game. In the textbook there aren't many problem sets, but while playing games on the AI system, I solved many questions so I could finish the quests and go to the next level."*

For students like Jake, the platform appeared to increase voluntary practice, which may contribute to improvements in mathematics performance.

#### *4.3.3. Seeking Greater Cognitive Challenge*

For higher-achieving students, the AI platform provided opportunities to engage with more challenging problems and progress at an individualized pace. In traditional classroom settings, these students often completed assigned tasks quickly and had to wait for peers to finish.

Cameron, a high-achieving student, described how the AI platform allowed him to continue learning without waiting for others:

*"Textbooks are too easy for me. When we solve textbook problems or worksheets, I usually finish in about five minutes. Then I have to wait for everyone else to finish. Sometimes I wait for twenty minutes. But when my teacher lets me use the AI tool, I can solve many complex problems on the system."*

This pattern suggests that AI-supported learning environments can provide enrichment opportunities for advanced learners by enabling continuous and self-paced problem solving.

#### ***4.3.4. Flexible Learning Opportunities Beyond Classroom Time***

Another observed pattern was the flexible use of the platform across different learning contexts. Some students continued using the AI system outside regular instructional time, such as during breaks or at home.

Abby, a student who required more time to complete learning tasks, appreciated the flexibility offered by the platform:

*"When activity time was short, I sometimes couldn't finish them. I liked this tool because I could also use it at home. My teacher let me use it during break time and at home."*

This flexibility appeared to benefit students who needed additional time for practice and task completion.

#### ***4.3.5. AI Disengagement***

Despite the positive experiences described above, not all students benefited from the AI platform. In particular, students with low digital literacy or negative attitudes toward technology sometimes struggled to engage with the system. These students were also identified as outliers in the quantitative analysis of mathematics achievement.

For example, David and Emma had difficulty using the platform due to limited experience with AI devices. During classroom observations, they frequently struggled to log into the system or navigate the interface. Their attitudes toward the platform were also negative.

David stated:

*"I hate computers. I don't like the AI platform at all. I hate math too."*

Similarly, Emma explained:

*"I don't know what to click to do what I want. I also forgot my password. I tried to remember it but couldn't. At that time I didn't do anything on the screen. I don't have much experience using computers. I don't even have my own phone."*

These cases illustrate that students' AI literacy and attitudes toward technology can influence their ability to benefit from AI-supported learning systems.

Overall, the qualitative findings suggest that students' interactions with the AI platform varied substantially depending on their prior achievement levels, learning preferences, and digital familiarity. While some students experienced increased confidence, engagement, and opportunities for advanced practice, others faced barriers related to technology use and motivation. These interaction patterns help contextualize the quantitative findings on achievement and interest trajectories by illustrating how students experienced the AI-based mathematics learning environment in practice.

# 5. Discussion

This study examined how students' mathematics achievement and interest evolved over time during an AI-based mathematics learning intervention. By combining linear mixed models, linear quantile mixed models, and qualitative classroom observations, the study provides a comprehensive perspective on both the overall trajectory and the heterogeneous learning experiences of students using an AI-supported learning platform. The findings reveal several key insights. First, students' mathematics achievement improved substantially during the intervention, and the improvement plateaued toward the end of the semester. Second, achievement trajectories differed across the performance distribution, with higher-achieving students showing earlier gains than lower-achieving students. Third, while students' interest in mathematics initially increased, the effect was not sustained throughout the semester. Qualitative findings further illustrate how different interaction patterns with the AI platform contributed to these outcomes.

## 5.1. Achievement Growth in AI-Supported Mathematics Learning

The results indicate that students' mathematics achievement increased significantly during the intervention. Both the descriptive statistics and the linear mixed model analyses showed a substantial improvement between the early and middle phases of the semester, particularly between Time 2 and Time 3. These findings are consistent with prior research suggesting that adaptive learning systems can support learning achievement by providing personalized practice opportunities and immediate feedback (du Boulay, 2019; Graesser et al., 2005; Holmes et al., 2018; Walkington & Bernacki, 2019).

AI-based learning environments can dynamically adjust task difficulty and provide individualized problem sequences based on learners' performance. Such adaptive mechanisms allow students to engage with problems that are appropriately challenging for their current level of understanding, thereby facilitating skill development and reducing learning gaps (Graesser et al., 2005; Huang et al., 2016). The qualitative findings reinforce this interpretation. For example, several students described how the AI system recommended problems that matched their skill level, enabling them to practice repeatedly until they achieved mastery.

However, the trajectory of achievement showed a noticeable plateau toward the end of the semester. Although scores at the final measurement remained significantly higher than the baseline, they were slightly lower than the peak observed in the middle of the intervention. One possible explanation is that the initial phase of AI-supported learning may generate rapid gains as students engage with the novelty of the platform and address foundational learning gaps. As the intervention progresses, however, tasks may become more cognitively demanding, leading to slower improvements in performance. This pattern supports that learning gains in technology-supported environments may follow a non-linear trajectory rather than a steady linear increase over time. These results suggest that AI-supported learning environments may produce rapid initial improvements but require additional instructional support to sustain long-term academic growth.

## 5.2. Heterogeneous Achievement Trajectories Across Performance Levels

A central contribution of this study lies in its examination of heterogeneous achievement trajectories using a distributional perspective. The LQMM results revealed that the benefits of the AI-based learning intervention were not uniformly distributed among students. Higher-achieving students demonstrated significant improvements earlier in the intervention, while lower-performing students required a longer duration to manifest significant improvements, with the intervention effects waning by the semester's end.

These findings resonate with the theoretical perspective of aptitude–treatment interaction, which suggests that instructional approaches may interact with learner characteristics such as prior knowledge and cognitive ability (Cronbach & Snow, 1981; Snow, 1991). In mathematics learning, prior knowledge plays a particularly important role because mathematical concepts are cumulative and hierarchical (Booth et al., 2014; Lee, 2024; Wilkins & Norton, 2018). Students with stronger prior knowledge may therefore leverage AI learning systems more effectively, as they are able to engage with more advanced problem-solving tasks from the outset. At the same time, the delayed improvement observed among lower-performing students suggests that AI-based learning environments may still support meaningful progress for students with weaker prior knowledge, albeit at a slower pace.

The qualitative findings offer further insight into these distributional patterns. Higher-achieving students reported that the AI platform allowed them to engage with more challenging problems and continue practicing without waiting for peers to complete classroom activities. In contrast, lower-achieving students often required additional time to build foundational understanding before they could benefit from more advanced tasks. For example, they described how the system initially provided easier problems, allowing them to experience early success before gradually progressing to more challenging tasks. This process of scaffolded progression may explain why learning gains among lower-performing students emerged later in the intervention.

Importantly, these findings illustrate why distributional approaches are necessary for evaluating AI-supported learning systems. Mean-based analyses may obscure heterogeneous effects across the performance distribution, potentially masking both the early advantages experienced by higher-achieving students and the delayed gains observed among lower-performing students. By examining learning trajectories across quantiles, the present study reveals that AI learning does not benefit all students equally. AI-supported learning environments may benefit students differently depending on their initial level of achievement.

### 5.3. Short-Term Engagement but Limited Sustainability of Mathematics Interest

While the intervention produced clear improvements in mathematics achievement, the results for mathematics interest revealed a more nuanced pattern. Students' interest increased during the early stage of the intervention but declined toward the end of the semester, ultimately returning close to baseline levels. This pattern suggests that the initial engagement generated by the AI platform may not have translated into sustained motivational changes. This result implies a need for motivational strategies to sustain for a longer period of time.

One possible explanation relates to the distinction between situational interest and individual interest. Situational interest can be triggered by novel or engaging learning environments, such as digital platforms with interactive features (Yeh et al., 2019). However, the development of more stable individual interest may require repeated experiences of competence and meaningful challenge over time. In the present study, the novelty of the AI platform, interactive features, and its gamified elements may have initially stimulated situational interest, but this effect diminished as students became accustomed to the AI learning environment.

The distributional analysis further revealed that the increase in interest was primarily observed among students in the middle and upper portions of the interest distribution. Students with initially low interest in mathematics did not demonstrate significant changes across the semester. This finding suggests that AI-supported learning systems may be more effective at amplifying existing interest rather than generating interest among students who are initially disengaged from the subject.

### 5.4. Interaction Patterns with the AI Platform

The qualitative findings provide important contextual insights into how students experienced the AI-based mathematics platform. Several distinct interaction patterns were identified, including small success experiences among lower-achieving students, game-based engagement among intermediate learners, and opportunities for advanced problem solving among higher-achieving students.

These patterns illustrate how adaptive learning environments can support personalized learning experiences. For example, struggling students reported that solving easier problems helped them gradually build confidence in mathematics. Intermediate students were particularly motivated by gamified features, which encouraged them to complete a large number of practice problems. High-achieving students valued the opportunity to work on challenging tasks and progress at their own pace.

However, the cognitive learning outcomes for the low-achieving student group showed significant variation within the group. While some low-achieving students experienced dramatic improvements in their mathematics achievement, others saw only minimal progress. Notably, the variability in performance among low-achieving students is much greater compared to that within high-achieving and intermediate groups. The qualitative data also revealed barriers to engagement for some students. In particular, students with low AI literacy or negative attitudes toward technology had difficulty interacting with the platform. These students were also identified as outliers in the quantitative analysis of achievement trajectories. Their experiences highlight an important limitation of AI-based learning systems (Lee & Brush, 2025). The effectiveness of AI may depend not only on the design of the platform but also on students' AI familiarity and attitudes toward digital learning.

## 6. Implications and Future Research

This study investigates the longitudinal trajectories of mathematics achievement and interest during an AI-supported mathematics learning intervention. The findings of this study contribute to the growing body of research on AI-supported education in several ways. The findings of this study have several implications for the design and implementation of AI-supported learning environments.

First, the heterogeneous achievement trajectories observed in this study suggest that adaptive systems should provide differentiated support for learners with varying levels of prior knowledge. Lower-achieving students may require additional scaffolding and extended opportunities to build foundational skills before they can fully benefit from adaptive learning environments.

Second, the results highlight the importance of sustaining students' motivation over time. While interactive interfaces may initially stimulate engagement, maintaining students' interest may require more meaningful learning experiences and additional instructional strategies that promote long-term perceptions of competence and autonomy.

Third, the presence of students who struggled to interact with the AI platform underscores the need to consider AI literacy when implementing AI-based learning tools in classrooms. Providing structured onboarding and technological support may help ensure that all students can effectively access and benefit from such systems.

Importantly, this study highlights that AI learning does not benefit all students equally. Additionally, this research underscores both the potential and the limitations of AI-supported learning environments in shaping cognitive and motivational outcomes.

While our study provides meaningful contributions, certain limitations should be acknowledged. First, the sample size was relatively small and drawn from a single classroom context, which may limit the generalizability of the results. Second, although the mixed-methods design provided valuable insights into students' learning experiences, future research could incorporate more systematic measures of students' interaction patterns with AI platforms using learning analytics data. Future studies may also explore how specific design features of AI-based learning systems influence both cognitive and motivational outcomes. For example, examining how adaptive feedback or recommendation algorithms influence student engagement could provide valuable insights for improving the design of educational AI systems.

**CONFLICT OF INTEREST STATEMENT**
The authors have no conflicts of interest to disclose.